\documentclass[conference]{IEEEtran}
\IEEEoverridecommandlockouts
\usepackage{cite}
\usepackage{amsmath,amssymb,amsfonts}
\usepackage{algorithmic}
\usepackage{algorithm}
\usepackage{graphicx}
\usepackage{textcomp}
\usepackage{xcolor}
\def\BibTeX{{\rm B\kern-.05em{\sc i\kern-.025em b}\kern-.08em
    T\kern-.1667em\lower.7ex\hbox{E}\kern-.125emX}}
\begin{document}

\title{Cantor Mapping Technique\\}

\author{
\IEEEauthorblockN{Kaustubh Joshi}
\IEEEauthorblockA{
kaustubhkj@gmail.com}
}

\maketitle

\begin{abstract}
A new technique specific to String ordering utilizing a method called "Cantor Mapping" is explained in this paper and used to perform string comparative sort in loglinear time while utilizing linear extra space.
\end{abstract}


\section{Introduction}
The whole process of comparative sorting is to compare two elements amongst each other to find out which one possess a higher and lower value (or equivalent). Due to this, the time complexity of each comparative sorting algorithm always depends on f(l) factor where l is the representation of the object being compared. Standard merge sort takes O(n*log(n)) time for numbers and O(n*l*log(n)) for strings. This f(l) factor is O(1) for numbers and O(l) for linear sequences. \\

Due to the nature of the f(l) factor, the time complexity of searching and sorting algorithms increases by a factor of f(l) depending on the object type. This is what causes string sorting to take longer when compared to standard numerical sorting. Non comparative methods overcome this by utilizing additional memory (which isn't always available to machines) as displayed by counting sort for a fixed upper and lower bound for integer values and burst sort which utilizes a stable trie structure.

\section{Problem Statement}
Given a random unsorted array of strings, what is the most optimum space and time complexity in which you can sort the result ? Is there a middle ground in case there isn't ?\\

Standard approach is to treat this as a problem in key based sorting and utilizing multi-array sort to solve for the same. This approach has a time complexity of O(n*$\bar{l}$*log(n)) and space complexity of O(1) and is multi valued sort extended to strings. \\
Another approach is to build a key index tree (or trie) and then perform Depth First Search in order to obtain the sorted values. This approach has a time complexity of O(n*$\bar{l}$) and space complexity of O(n*l) and is called as burst sort\cite{b5}. \\

The value compare approach is preferred when memory is of the highest importance and the system is relaxed on running time. The burst sort approach is preferred when the system has a sufficient amount of memory and requires fast execution. \\
There is a trade off for choosing either of the two, however using the given method of Cantor Mapping, we wish to propose a middle ground with O(n) of memory space and O(n*log(n)) of time complexity.

\section{Cantor Mapping}
The idea of Cantor Mapping can be considered as the fusion of Cantor Sets and closed space hashing. A Cantor Set is a closed set with remarkable properties such as nowhere dense, perfect set, low capacity dimension, recursively defined, etc. \\
Length of each segment at depth d = $l_{d}$ = $(1/3)^{d}$ \\
Number of segments at depth d = $n_{d}$ = $(2)^{d}$ \\
Total length for given depth = $L_{d}$ = $n_{d} * l_{d}$ \\

\begin{figure}[htbp]
\centerline{\includegraphics[width=9cm,height=10cm,keepaspectratio]{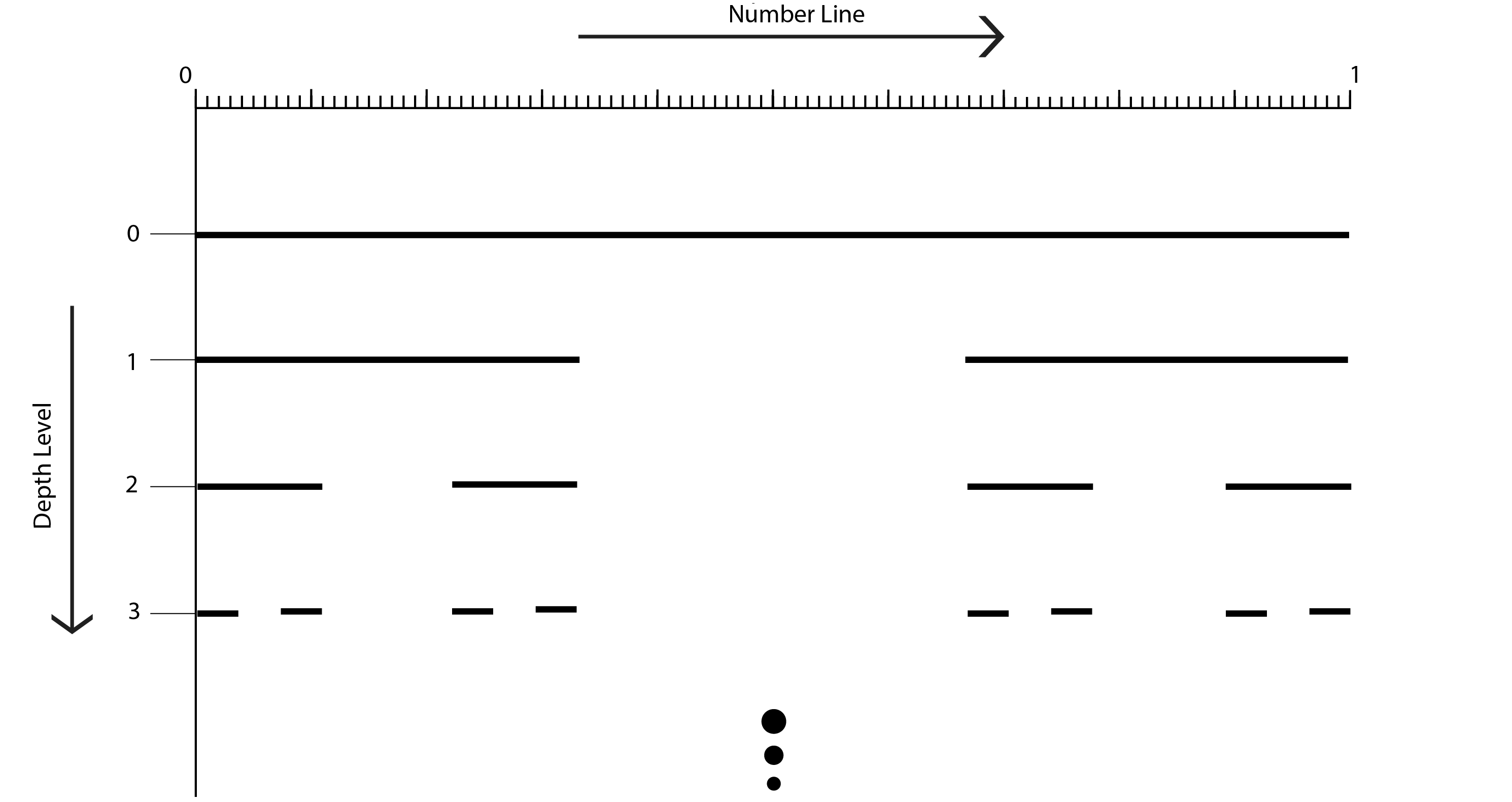}}
\caption{A Cantor Set}
\label{cantorSet}
\end{figure}

Hashing is a method that is used for assigning values to objects using the modular polynomial ring method to ensure the co-domain is bounded. This technique is used in maps to achieve O(1) time lookups and in rabin-karp\cite{b1} for single valued string search\\

Cantor Mapping is a technique used to ensure that a string value is provided with an equivalent numerical value. The unique property of this mapping is that any string can be used for searching and ordering based on this numerical value rather than having to iterate and compare its elements. \\
Unsigned floating point numbers are assigned monotonically, hence an ordering of the number ensures that the corresponding index based string value is also ordered. While this method is analogous to Polynomial Hashing, it should not be used to replace it because unlike hashing, mapping is injective by nature where as hashing is not injective by nature. \\

\subsection{Technique}
We need a function f(x): R $\rightarrow$ N for performing ordinal ranking.  Our function needs to be strictly monotonic in nature\cite{b7}. I.e. given $f(x1) < f(x2) \implies x1 < x2$. \\
We could use an intermediate function for increasing/decreasing number spacing. A function m(x): $R \rightarrow R$ such as the exponential function,linear functions, polynomial functions, etc. are examples of what can be utilized. Weakly monotonic functions cannot be utilized as this will lead to ordinal ranking violation when performing f(x).\\
We also need a function g(z): $String \rightarrow N$, such that $z \in \{Z* \cup T\}$  (where T is a terminal and Z is a set of non terminals) which acts as a series converter of sorts. The main purpose of this function is to reduce the length of the data sequence\\
Over here we utilize the composite function $f \circ m \circ g$ which is specified as follows:\\
T(x): String $\rightarrow$ N where T(x) = $\sum_{i=0}^{|S| - 1}$ $s_{i}/x^{i}$ \\
The value of x should be = $|T| + \epsilon$ where $\epsilon \ge 1$. For this paper the total alphabet size was 26 and $\epsilon = 4$ \\

\begin{algorithm}
\caption{Cantor Mapping}
\begin{algorithmic}
\STATE $S \leftarrow$ A string
\STATE $M \leftarrow$ A character to number order Map
\STATE $\epsilon := 4$
\STATE $x := |T| + \epsilon $
\STATE $Result := 0 $
\FOR{$(i = |S| - 1; i \ge 0; i--)$}
\STATE {$Result = ((Result/x) + M.get(S[i]))$}
\ENDFOR
\STATE return Result
\end{algorithmic}
\end{algorithm}

The proof of the function being monotonic (given proper value of x is as follows):\\
Assume two strings B and C (where B is dictionary ordered before C) which have a common prefix of size l. \\
$\implies$ T(B[0,l-1]) = T(C[0,l-1]) and $B[l] \neq C[l]$\\

As per the procedure of cantor mapping:\\
T(B[l,$|B| - 1$]) = $\sum_{i=l}^{|B|-1}$ $B[i]/x^{i}$ \\
T(C[l,$|C| - 1$]) = $\sum_{i=l}^{|C|-1}$ $C[i]/x^{i}$ \\

Since B should come before C:\\
$T(C[l]) > T((B[l,|B| - 1])$ \\
$\implies C[l]/x^{l} > (B[l]/x^{l} + B[l+1]/x^{l+1} + \ldots)$
$\implies (C[l] - B[l])/x^{l} > B[l+1]/x^{l+1} + B[l+2]/x^{l+2} \ldots$\\

As each character is bounded by the maximum attainable value $\zeta$  (where $\zeta$ = value of the last non-terminal)\\

$\implies (C[l] - B[l])/x^{l} > \zeta/x^{l+1} + \zeta/x^{l+2} \ldots  \\= (\zeta/x^{l+1})(1/(1-(1/x)))$\\
$\implies (C[l] - B[l]) > (\zeta/x)* (x/(x-1))$\\
$\implies x > \zeta /(C[l] - B[l]) + 1$\\

Hence maximizing the above equation, we can get the correct value of x in order to ensure monotonicity.

\section{Utility}
Here three major types of sorting methods relevant to string operations have been mentioned for exemplification purposes. They are as follows:

\subsection{Standard String Sorting}
Let A be a collection of strings that we wish to sort in lexicographical order, using the Cantor Mapping technique we first preprocess and store every string based on its respective index. This task happens in exact n*$\bar{l}$ time and requires O(n) space. Once this is completed we can utilize a standard comparison sort which compares the unsigned float values and returns the index ordered array. This process happens in O(n*log(n)) time. \\
Hence the total time complexity can be seen as O(n*$\bar{l}$ + n*log(n)). $\bar{l}$ and log(n) will be the competing factor behind which time complexity the algorithm runs in. ($\bar{l}$ is the average word length and n is the number of strings present in the collection)

\begin{figure}[htbp]
\centerline{\includegraphics[width=8cm,height=10cm]{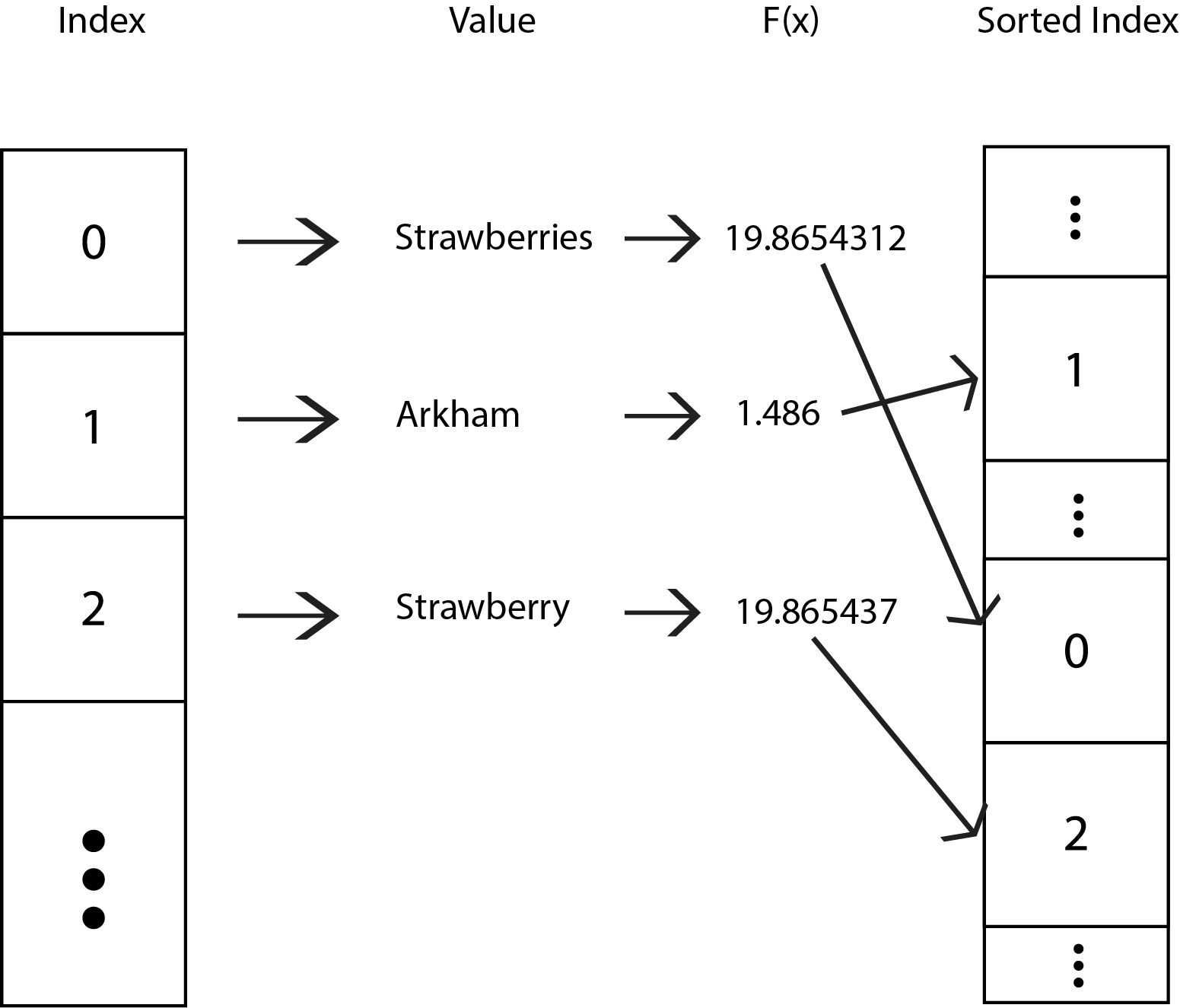}}
\caption{Standard String Sorting}
\label{SSS}
\end{figure}

While floating number comparison is more complex compared to standard integers, the run-time of floating number comparison is $\Omega(integer comparison)$. Hardware optimizations are possible using a Floating Point Unit which can reduce the constant factor even more. Software optimized algorithms can be used too\cite{b6}\\

\subsection{Split-wise String Sorting}
Let A be a collection of strings that we wish to sort in lexicographical order and assuming there exists a system,problem or hardware restriction for computing precise floating point numbers, we can create chunks of certain size and perform standard multi-array sort for the same. \\

This sorting requires O(n* max($l_{i}$) * log(n) / k) time  and O(n* max($l_{i}$) / k) space (where k is the split size and $l_{i}$ is the size of the $i^{th}$ string). Worst case scenario occurs where the value of k = 1 as this will correspond to the standard comparative string sorting\cite{b3}.\\

\subsection{Suffix String Sorting}
Let S be the string which we wish to create a Suffix Array from. We can preprocess the initial index based values for all indexes in total O(n) time. \\
Following this we can subject the array to comparison sort leading us to suffix sorted results in O(l*log(l)) time. While this is comparable to older suffix sort algorithms\cite{b2}, newer ones\cite{b4} can perform sorting in O(l) time and hence isn't a viable alternative to modern suffix sorting algorithms\\

\begin{algorithm}
\caption{Suffix pre-processing in O(n) time}
\begin{algorithmic}
\STATE cantorMap = Function calculates cantor mapping
\STATE $S \leftarrow$ String we wish to find suffix array of
\STATE A[] := Array of size $|S|$
\STATE $\epsilon := 4$
\STATE $x := |T| + \epsilon $
\STATE $currentVal := 0 $
\FOR{$i = 0; i < |S|; i++$}
\STATE {$currentVal = ((currentVal/x) + cantorMap(S[|S| - 1 - i])$}
\STATE $A[i] = currentVal$
\ENDFOR
\end{algorithmic}
\end{algorithm}

\subsection{Calculation optimizations}
In the case of frequently occurring prefixes, or split wise string sorting, we utilize a hash table and then perform remaining calculation. Assuming a certain prefix exists and the string S = prefix + remainder. The result will be (hashTable[prefix] + cantorMap(remainder)/$x^{|prefix|}$)

\section{Conclusion}
The Cantor Mapping technique was explained along with the  math properties which must be used in order to utilize faster string value comparisons using unsigned floating value conversion.\\

Use cases help illustrate how to utilize this technique properly for the given sorting type. Ultimately using software/hardware floating point comparison string comparative sorting is achieved in O(n*log(n)) time while using only O(n) floating value space.\\

\end{document}